\documentclass[newstyle,twocolumn,proceedings]{rmaa}
\newcommand{\lya}{Ly$\alpha$}
\newcommand{\ha}{H$\alpha$}
\usepackage{rmaacite}
\title{Observations of radio galaxy MRC~1138-262:\\
merging galaxies embedded in a giant \lya\ halo}
\author{J. D. Kurk, H. J. A. R\"ottgering and G. K. Miley
  \affil{Sterrewacht Leiden, The Netherlands}
 \and L. Pentericci 
  \affil{Max Planck Institut f\"ur Astronomie, Heidelberg, Germany}}
\fulladdresses{
\item Kurk, R\"ottgering and Miley: Sterrewacht Leiden,
P.O. Box 9513, 2300 RA, Leiden, The Netherlands.
\item Pentericci: Max Planck Institut f\"ur Astronomie, K\"onigstuhl 17, 
D-69117, Heidelberg, Germany. }
\shortauthor{Kurk et al.}
\shorttitle{Observations of radio galaxy MRC~1138-262}

\abstract{The radio galaxy MRC~1138-262 at $z=2.16$ is most likely a
brightest cluster galaxy in an early stage of evolution. Here we
present observations of the luminous emission line halo and the
stellar components of this radio galaxy. Optical narrow band imaging
shows a very extended ($\sim 160$ kpc) and luminous \lya\
halo. Infrared narrow band imaging reveals a much smaller \ha\ halo
with a morphology very different from that of the \lya\ halo. We
advocate a model in which the inner part of the halo is photoionized
by direct AGN illumination or by UV photons from young stars. Far from
the nucleus ($\sim$ 25 kpc), there is a region of greatly enhanced
\lya\ emission. At this location, it is likely that shock ionization
is important as indicated by a bend in the radio jet. Spectroscopy of
several continuum clumps in the halo shows that, although there are
striking differences between the emission and absorption features of
the spectra of various regions, they have properties similar to those
of Lyman-break galaxies. This is further evidence for a scenario in
which massive galaxies form hierarchically from smaller building
blocks.}

\listofauthors{J.~D.~Kurk \& L.~Pentericci \& H.~J.~A. R\"ottgering \& 
G.~K.~Miley}
\indexauthor{Kurk, J.~D.}
\indexauthor{Penterricci, L.}
\indexauthor{R\"ottgering, H.~J.~A.}
\indexauthor{Miley, G.~K.}

\begin{document}
\maketitle

\section{Introduction}
\label{sec:intro}

In this paper we present observations which were obtained as part of
our campaign to search for clusters around high redshift radio
galaxies (HzRGs). There are several indications (e.g.\ Pentericci et
al.\ 1999) \nocite{pen99} that powerful HzRGs tend to be in the center
of forming clusters. MRC~1138-262 at redshift 2.16 is one of the best
candidates for such a search. The indications for 1138-262 being at
the center of a cluster include (a) the very clumpy morphology as
observed by the HST \cite{pen98}, reminiscent of model predictions for
a massive merging system \cite{car94}; (b) the extremely distorted
radio morphology and largest radio rotation measure (6200 rad
m$^{-2}$) in a sample of more than 70 HzRGs, indicating that 1138-262
is surrounded by a hot, clumpy and dense magnetized medium
\cite{car97}.  On the basis of these arguments we targetted 1138-262
for a cluster search using narrow band imaging and multi object
spectroscopy. These observations resulted in the discovery of a
structure of 14 \lya\ emitting galaxies in the region of this radio
galaxy \cite{pen00}. Here we present additional results from these
observations that shed light on the origin of the emission line gas,
the ionization processes and the constituents of the presumably young
galaxy.

\begin{table}
\begin{center}
\caption[]{Observation log}
\begin{tabular}{l l l c}
\hline \hline     
Technique & Date & Config & Time \\
    &     &          & (hour) \\
\hline
Optical   & April 1999   & Bessel B & 0.5 \\
Imaging   &              & O{\small II}/3814 & 4.0 \\
\hline
Optical   & March 2000   & Mask A   & 5.5 \\
\multicolumn{2}{l}{Spectroscopy} & Mask B   & 6.0 \\
\hline
Infrared  & March 1999   & K$_s$    & 1.5 \\
Imaging   & March 2000   & 2.07$\mu$& 6.0 \\
\hline\hline
\label{obs}
\end{tabular}
\end{center}
\vspace{-0.8cm}
\end{table}

\section{Observations and results}
\label{sec:obs}

Using optical broad and narrow band images, obtained at the 8.2m ESO
VLT\footnote{This research is based on observations carried out at the
European Southern Observatory, Paranal, Chile.} Antu telescope with
FORS1, we found 50 objects that were candidates for galaxies with
large \lya\ equivalent width at $z$ = 2.16 \cite{kur00}. Multi object
spectroscopy with the same instrument gave redshifts for 15 candidates
\cite{pen00}. We have also obtained infrared broad and narrow band
images with ISAAC at the Antu telescope to search for \ha\ emitting
companions of the radio galaxy.  Table~\ref{obs} summarizes the dates
and integration times of these observations. The optical and infrared
observations are also very suitable to study MRC~1138-262 itself: from
the optical imaging we have obtained a detailed map of the \lya\ halo,
from the spectroscopy deep spectra of several components of the galaxy
and from the infrared imaging a map of the \ha\ halo.

\begin{figure}
  \includegraphics[angle=90,width=\columnwidth]{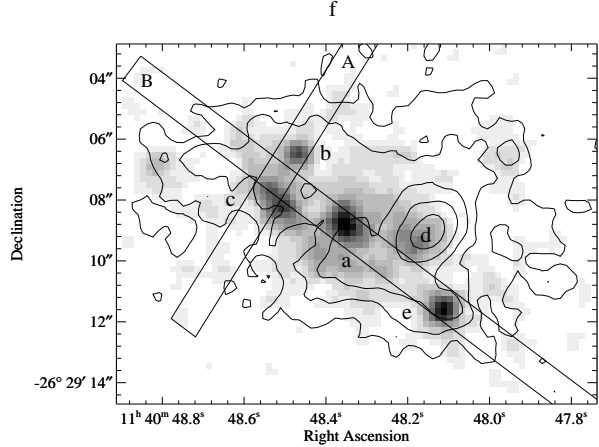}
  \vspace{-0.9cm} \caption{B band image of 1138-262 (greyscale)
  overlayed with contours of Ly$\alpha$ emission. Also shown are the
  locations of two slits used for spectroscopy, labeled with upper
  case letters. Lower case letters label the various parts of the
  radio galaxy, as in Pentericci et al.\ (1997). \label{blya}}
  \vspace{-0.4cm}
\end{figure}

\subsection{Optical imaging}
Image reduction resulted in 6\farcm8$\times$6\farcm8 narrow and broad
band images with seeing 0\farcs9 and 1$\sigma$ limiting AB magnitude
per square arcsecond of 27.8 and 28.1 respectively.

The B band image (see greyscale in Fig.~\ref{blya}) shows that
MRC~1138-262 has a very clumpy morphology, quite unlike modern-day
massive ellipticals. Instead, there are 7 seperate barely resolved
knots in a 15\arcsec$\times$10\arcsec\ region, some aligned with the
direction of the radio jet axis. In Fig.~\ref{blya} these components
are labeled by lowercase letters. \nocite{pen97} Component {\sl a} is
identified with the nucleus, hosting the AGN of the radio galaxy.

We have obtained a \lya\ emission image of the radio galaxy from the
difference between the narrow and the broad band image. The \lya\
emission map (contours in Fig.~\ref{blya}) strikingly shows the
enormous size of the halo encompassing all the continuum clumps. Most
continuum clumps have a counterpart in the \lya\ map, but the peak of
the \lya\ emission does not coincide with the most luminous continuum
source ({\sl a}), but with a much fainter one ({\sl d}).

\subsection{Infrared imaging}

The reduced 2\farcm5$\times$2\farcm5 narrow and broad band images have
an average seeing of 0\farcs5 and 1$\sigma$ limiting AB magnitude per
square arcsecond of 23.9 and 24.0 respectively.

Contrary to the B band image, the K$_s$ band image shows one main
bright unresolved component, accompanied by two small clumps in the
North East direction. This structure coincides with the nucleus and
components {\sl b} and {\sl c} in the optical image.

From the narrow and broad band images we have constructed an \ha\
emission image in the same way as the optical \lya\ emission
image. The \ha\ morphology is very different from the \lya\
morphology. The \ha\ emission is dominated by the bright unresolved
nucleus. It has a westward extension of about 5\arcsec, which we
associate with component {\sl d}. There are three more faint \ha\
emission regions, of which two are coinciding with a local enhancement
of \lya\ emission.

\begin{figure}
  \includegraphics[width=\columnwidth]{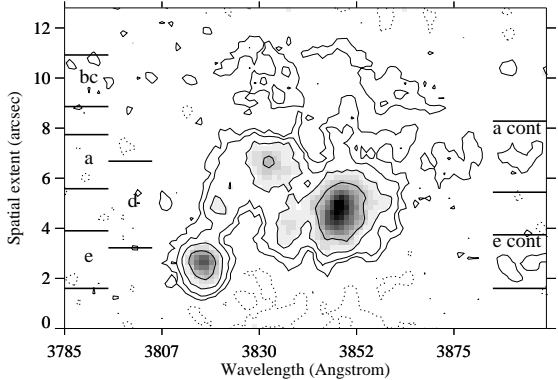}
  \vspace{-0.7cm}
  \caption{Contour plot of the \lya\ region of the slit in mask
         B. Aperture limits are indicated by thick straight lines for
         the individual components denoted by their names inside the
         apertures. \label{b_aps}}
  \vspace{-0.3cm}
\end{figure}

\subsection{Optical spectroscopy}

\begin{figure}
  \includegraphics[width=\columnwidth]{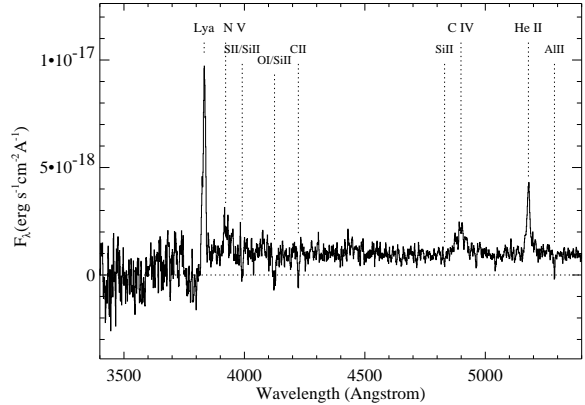} \vspace{-0.7cm}
  \caption{Spectrum of component {\sl a} in mask B. Emission lines are
  indicated. The continuum shows several interstellar absorption
  lines. \label{B10_at}}
  \vspace{-0.3cm}
\end{figure}

We used several FORS1 masks to observe the candidate \lya\ emitters
near MRC~1138-262 and in two masks we employed one central slit to
observe several clumps of the radio galaxy. The positions of these
central slits are shown in Fig.~\ref{blya}. The spectra cover a
wavelength range from approximately 3180 to 5550 \AA\, and the
1\arcsec\ slitwidths gave a resolution of $\sim$ 5 \AA\ corresponding
to about 400 km s$^{-1}$. The slits in mask A and B each cover at
least three components in the halo of the radio
galaxy. Fig.~\ref{b_aps} shows the complex intertwined two-dimensional
spectrum of the \lya\ lines from the components in the slit of mask
B. All components observed show \lya\ emission, but components {\sl a}
and {\sl b} also have continuum emission with interstellar absorption
lines. Components {\sl c} and {\sl e} do show a continuum but no
absorption lines are detected, despite the good s/n. Component {\sl a}
(see Fig.~\ref{B10_at} for its one-dimensional spectrum) is the only
one possessing metal emission lines.

\begin{figure*}
\begin{center}
  \includegraphics[angle=90,width=14.2cm]{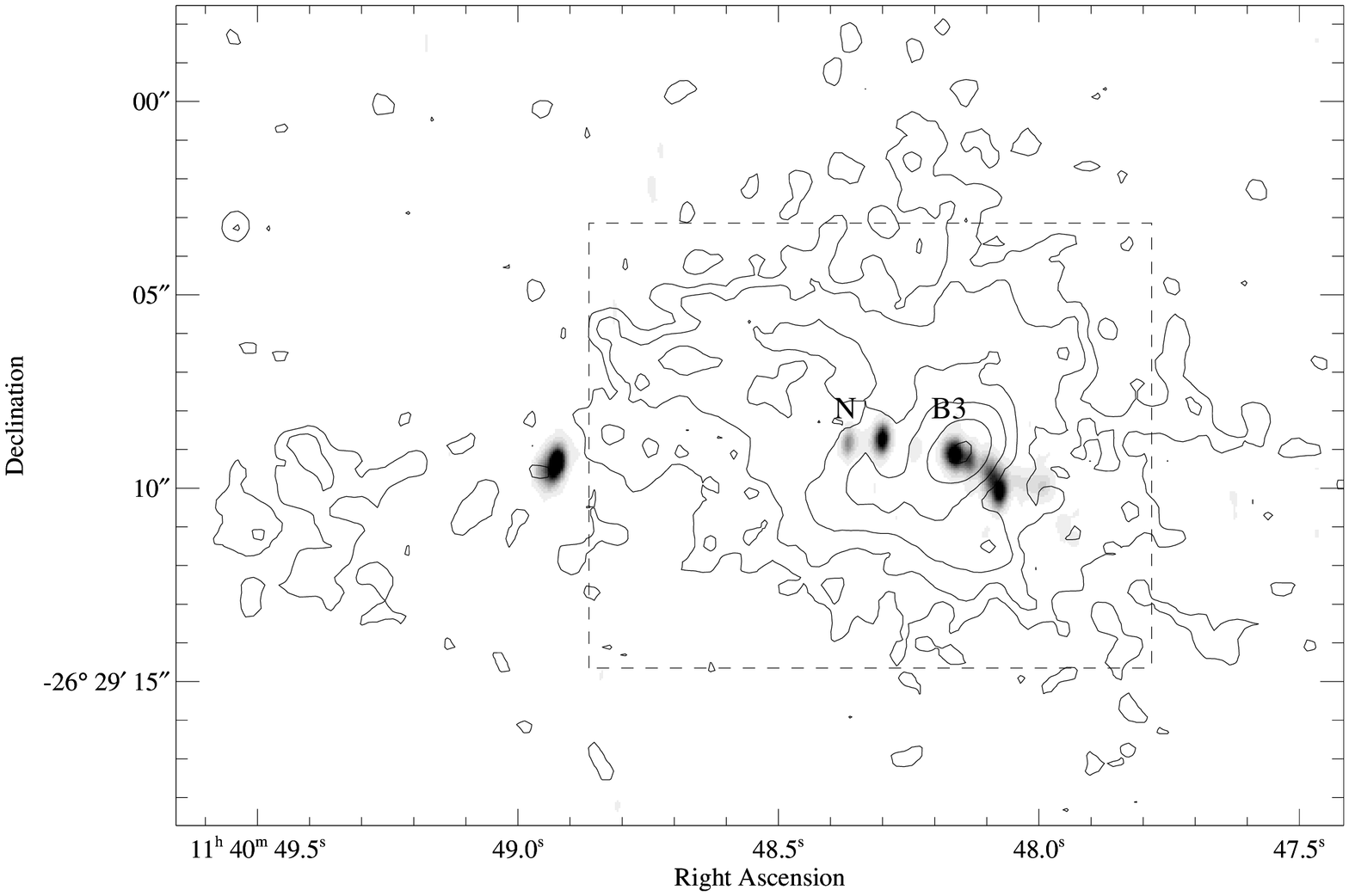} \vspace{-1.0cm}
  \caption{Contour plot of the \lya\ emission with radio continuum
  emission at 8.1 GHz superimposed in greyscale. Contours are a
  geometric progression in steps of 2, starting from 2 times the
  background rms noise. The nucleus (N) and the radio knot where the
  jet bends (B3) are indicated, following Pentericci et al.\
  (1997). The rectangle is enlarged in Fig.~\ref{ratio}.
  \label{xlya}} \vspace{0.1cm}
\end{center}

\begin{center}
  \includegraphics[angle=90,width=14.2cm]{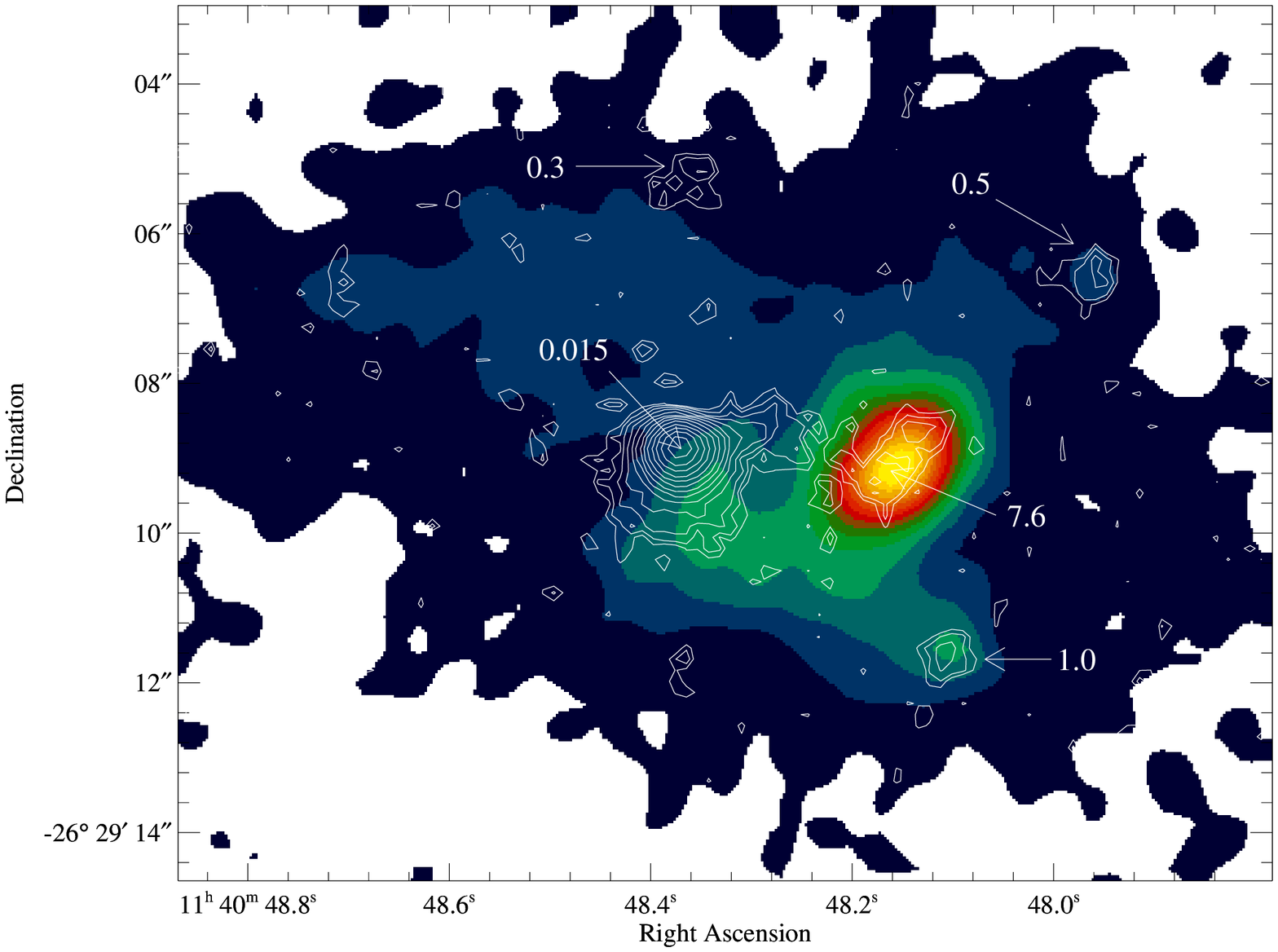} \vspace{-1.0cm}
  \caption{Close-up of the boxed region in Fig.~\ref{xlya}. The \ha\
  emission is represented by contours, which are geometrically
  progressing in steps of 2$^{1\over 2}$, starting from 2 times the
  background rms noise. The underlying greyscale represents the \lya\
  emission. \lya/\ha\ ratios are indicated: the minimum of 0.015 at
  the nucleus and the maximum of 7.6 at the location of the bend in
  the radio jet, average values of 0.3, 0.5 and 1.0 for three presumed
  starburst regions.\label{ratio}}
\end{center}
\end{figure*}

\section{Discussion}
\label{sec:dis}

The fact that \lya\ and \ha\ emission are radiated by the same source
but transported in a different way -- \lya\ is a resonant line, while
\ha\ is not -- gives us the opportunity to study the combined presence
of neutral hydrogen as a scattering medium and dust as an absorbing
medium. In most astrophysical cases where hydrogen is photo-ionized,
the transition from n=2 to 1 is believed to be optically thick (Case B
recombination, Osterbrock 1989\nocite{ost89}). This dictates a
\lya/\ha\ ratio in the range of 8 to 12 depending on temperature and
density of the gas. The \lya/\ha\ ratio we detect in the halo of
MRC~1138-262 varies with position, but is always smaller than 8 (see
Fig.~\ref{ratio}).

It is striking that there is only weak \lya\ present at the position
of the strongest \ha\ clump with an observed \lya/\ha\ ratio of $\sim
0.015$. The presence of strong continuum emission here suggests
ionization by UV radiation from hot stars or alternatively from the
nearby AGN. The \lya\ photons are resonantly scattered by large
amounts of neutral hydrogen and possibly absorbed by dust, while the
\ha\ photons are relatively unaffected by these two processes.

The second region of interest is the \lya\ peak, which is situated at
the position where the radio jet has been bent, as shown in
Fig.~\ref{xlya}. We propose as the likeliest explanation that the
radio jet encounters a relatively dense medium at this
location. Propagation through the dense clump changes the jet
direction while driving shocks into the medium thereby ionizing
hydrogen gas. This behaviour is reminiscent of that in 3C277.3
\cite{bre85} where the radio jet is observed to bend at the position
of highest ionization gas clump. The \lya/\ha\ ratio in this region is
on average about 4 and at its peak at a value of 7.6, close to the
expected value for Case B recombination. Since this is far from any
continuum clump, it is not surprising that less neutral hydrogen
and/or dust is present.

Apart from three smaller regions with mean \lya/\ha\ ratios of 0.6,
\ha\ is not detected anywhere else within the \lya\ structure, leading
to lower limit \lya/\ha\ ratios from 0.1 to 2.0. A significant part of
the \lya\ emission in these regions is probably scattered emission and
no \ha\ is expected here. The \lya/\ha\ ratio of all emission measured
is 2.0. A correction for galactic extinction raises this ratio by
about 15\%. McCarthy et al.\ (1992)\nocite{mcc92} determined \lya/\ha\
ratios for two radio galaxies at $z=2.4$ by spectroscopy and they find
extinction corrected values of 4.7 and 3.5, slightly higher than our
value of 2.3. We conclude that, although the \lya/\ha\ ratio in the
powerful radio galaxy MRC~1138-262 has extremely low values in one
place due to scattering and very high values in another due to the
absence of a scattering medium, the average ratio is still a factor
$\sim 3$ below the recombination value, implying that at least some
dust is present.

The nucleus is the sole component observed which exhibits the metal
emission lines \ion{N}{5}, \ion{C}{4} and \ion{He}{2}. Using the
ratios of these lines as metallicity indicator (Vernet et al.\
2000)\nocite{ver00}, we derive that this clump's metallicity is three
times solar. From the absence of detectable metal emission lines in
other clumps with strong \lya\ emission, we surmise that there are
differences in metallicity among the clumps, consistent with the
metallicity gradient measured in MRC~2104-242 (Overzier et al.\ in
these proceedings).

The spectra of two continuum components show absorption lines due to
interstellar metal elements. The absorption lines in the central clump
are redshifted with respect to the \lya\ line by $\sim 1100$ km
s$^{-1}$. This is different from what is observed in the HzRG 4C41.17,
where the absorption is blueshifted with respect to the emission
\cite{dey97}. The equivalent widths of these lines are comparable with
the lines in the restframe UV spectrum of MS~1512-cB58 \cite{pet00}
and an average spectrum of 11 high-redshift galaxies in the Hubble
Deep Field \cite{low97}.

\section{Conclusion}
\label{sec:con}

Observations of the giant ionized gaseous halo which encompasses the
continuum clumps of 1138-262 suggest that hydrogen in this system is
ionized by two main sources. In the nucleus, the UV radiation
originates either in the AGN itself or in massive young stars. Near
the bend in the string of radio emission knots, shocks in the ambient
medium caused by the radio jet can explain the origin of the ionizing
radiation. Apart from small starburst regions which emit some \ha\
further away from the nucleus, \ha\ is not detected in the large \lya\
halo.  We conclude that the \lya\ emission observed far from any
continuum sources is scattered radiation and that there is dust
present in the radio galaxy which absorbs some of the scattered \lya\
photons.\adjustfinalcols The high metallicity derived from the
emission line ratios in the spectrum of the nucleus supports this
conclusion.

The characteristics of the interstellar absorption lines found in two
continuum components are similar to the absorption lines measured in
the restframe UV spectra of Lyman-break galaxies (LBGs). We propose,
therefore, that the components in the halo of 1138-262 are small
galaxies in various stages of evolution, some resembling LBGs, which
are falling in into the potential well of this system. Such a scenario
would agree with the simulations of hierarchical formation models for
massive galaxies (e.g.\ White et al.\ 1991).\nocite{whi91}
Extrapolating on the basis of these models from $z$ = 2.2 to the
present, 1138-262 would be transformed into a tranquil but bright
elliptical galaxy surrounded by its cluster companions.

The new images and spectra obtained with the VLT show that the complex
system of stars, gas and ionizing radiation, which we call
MRC~1138-262, contains significant amounts of neutral hydrogen gas,
but has also stellar components with properties similar to that of
LBGs. This is further evidence that MRC~1138-262 at $z$ = 2.16 is a
galaxy in the process of formation by hierarchical merging of LBGs.


\end{document}